\documentclass[aps,pre,twocolumn,showpacs,floatfix,superscriptaddress]{revtex4}
\usepackage{graphicx}
\usepackage{bm}
\usepackage{amssymb}
\bibstyle{apsrev.bib}

\begin{document}
\title{Universality class of the pair contact process with diffusion}
\author{F. Smallenburg}
\affiliation{Institute for Theoretical Physics, Utrecht
University, Leuvenlaan 4, 3584 CE Utrecht, The Netherlands}
\author{G. T. Barkema}
\affiliation{Institute for Theoretical Physics, Utrecht
University, Leuvenlaan 4, 3584 CE Utrecht, The Netherlands}
\affiliation{Institute-Lorentz, University of Leiden,\\Niels
Bohrweg 2, 2333 CA Leiden, The Netherlands}

\date{\today}

\begin{abstract}
The pair contact process with diffusion (PCPD) is studied with a
standard Monte Carlo approach and with simulations at fixed
densities. A standard analysis of the simulation results, based on
the particle densities or on the pair densities, yields
inconsistent estimates for the critical exponents.  However, if a
well-chosen linear combination of the particle and pair densities
is used, leading corrections can be suppressed, and consistent
estimates for the independent critical exponents
$\delta=0.16(2)$, $\beta=0.28(2)$ and $z=1.58$ are obtained. Since
these estimates are also consistent with their values in directed
percolation (DP), we conclude that PCPD falls in the same
universality class as DP.
\end{abstract}

\pacs{05.70.Ln, 05.50.+q, 64.60.Ht}

\maketitle

\section{Introduction}

In the fermionic one-dimensional pair contact process with diffusion
(PCPD), the model studied here, point particles can diffuse on a line,
and reactions can occur when two particles end up next to each other. The
particles can then both be annihilated, or if there is a free site next
to the pair, a new particle can be created.  The reactions present in
the PCPD model are:
\begin{eqnarray}
\begin{array}{ccc}
\left\{
\begin{array}{ccc}
AA0 & \rightarrow & AAA \\
0AA & \rightarrow & AAA
\end{array} \right.
&{\rm each\,\,with\,\, rate}&
\frac{(1-p)(1-d)} 2 \label{2Ato3A} \\
AA~ \rightarrow ~ 00
&{\rm with \,\,rate}& p\,(1-d) \label{2Ato0}\\
A0  \leftrightarrow  0A &{\rm with \,\,rate}&  d \label{A0to0A}
\end{array}
\label{eqrates}
\end{eqnarray}
with $0 < d < 1$ and $0 < p <1$.

As for other systems of this kind, the scaling relations which are
expected to hold for the density $\rho$ and correlation length
$\xi$ are
\begin{eqnarray}
\begin{array}{rclc}
\rho_{p=p_c} & \sim & t^{-\delta} & (\epsilon = 0) \\
\rho_{t\to \infty} & \sim & \epsilon^\beta  & (\epsilon > 0)\\
\xi & \sim & \epsilon^{-\nu_{\perp}} &(\epsilon \neq 0)\\
\xi & \sim & t^{1/z} & (\epsilon = 0)
\end{array}
\end{eqnarray}
in which $\epsilon = p_c - p$ is the distance to the critical point. The
critical exponents in these scaling relations define the universality
class a system belongs to\cite{hinr00,marr96}.
Two firmly  established universality classes so far for systems of this kind
are the Directed Percolation (DP) and Parity Conserving (PC) classes.

Since the introduction of PCPD a decade ago \cite{howard97}, or at least
a model closely resembling it, it has
attracted much attention. The main reason for this attention is that,
while its symmetries and conservation laws are seemingly identical
to directed percolation (DP), numerical estimates of its critical
exponents seem to place it in a different universality class. It has
been conjectured before by Grassberger \cite{grass82} and Janssen
\cite{jans81} that critical points with a unique absorbing state and
a single order parameter will fall into the DP universality class,
making this a likely candidate for PCPD as well. However, the fact that
the PCPD absorbing state is not unique makes this less certain, and it
has been suggested by earlier studies that it might not be possible
to describe PCPD with a single order parameter \cite{jans04,park05}.
Table \ref{tab01} shows values reported for the critical exponents of PCPD
from previous studies, as well as the accurately known values for these
exponents in the DP universality class. So far, while it is obvious much
research has been done on this subject, there is still a certain amount
of disagreement in the results.  Generally, the differences between the
measured exponents and the values for directed percolation have been
significant. It has been argued, however, that the discrepancy in $\delta$
between PCPD and DP is due to severe finite-size and finite-time effects~\cite{bark03},
and recent ultralong simulations~\cite{hinr06} show a clear
trend of $\delta$ towards its DP-value.

Other recent studies have looked at field theory approaches for PCPD
\cite{jans04}, and looked at a bosonic version of the model, where
multiple particles can exist in one spot \cite{kock03,hinr06}. Another
study \cite{kwon07} has investigated the structure and behavior
of clusters in PCPD, to clarify the slow approach of PCPD to its
asymptotic scaling regime.  Investigation of the crossover from PCPD to
DP \cite{park06} has yielded evidence towards non-DP scaling.

At this point, it is still unclear what universality class PCPD
does belong to. Most studies so far conclude that PCPD might belong
to a new univerality class \cite{kock03, park06}, or even that it
might belong to several ones based on the value of the parameter $d$
\cite{odor00,dick02,park02,odor03}. On the other hand, in a recent study
by H. Hinrichsen \cite{hinr06}, the critical exponent $\delta$ was shown
to display a significant drift towards the DP value, providing evidence
towards a single universality class. However, this tendency has not been
clearly shown for the other two exponents so far.

A previous study (Ref. \cite{bark03}) has provided numerical evidence
that, at the critical point, the ratio between the pair and particle
densities tends to a non-zero, finite value when the simulation time
tends to infinity. Also the high-quality data of Ref.~\cite{hinr06}
confirms the convergence to a nonzero ratio. A visual inspection of
the system shows big clusters, separated by increasingly large empty
(or near-empty) regions.  Since these clusters tend to have a finite
density in particles as well as in pairs, the convergence of their ratio
to a nonzero value is not unexpected.

Based on this convergence of the ratio of particles and pairs, we will
show in this paper that PCPD has strong correction terms in its scaling
relations when it has not yet reached the thermodynamical limit. These
correction terms will distort any direct measurements of the exponent.
The correction terms for these quantities, however, do not turn out
to be exactly the same.  We will show that if both the particle- and
the pair-densities are combined in the analysis, these corrections can
in some cases be suppressed, allowing a more accurate result for the
exponents. The resulting estimates for the critical exponents of the
PCPD model are consistent with the DP universality class.

\begin{table*}[btp]
\begin{tabular}{ | l || *{6}{@{\hspace{1mm}}l @{\hspace{3 mm}}|}}
\hline
Study & Year & $d$ &$\delta$ & $\beta$ & $z$ & $\beta/\nu_\perp$ \\
\hline \'Odor \cite{odor00} & 2000 &  0.1 & 0.275(4) & 0.58(1) & - & - \\
  &  & 0.5 & 0.21(1) & 0.40(2) & - & - \\
  &  & 0.9 & 0.20(1) & 0.39(2) & - & - \\
\hline
Carlon, Henkel & 2001 & 0.1 & - & - & 1.87(3) & 0.50(3)\\
and Schollw\"ock \cite{carl01} & & 0.5 & - & - & 1.70(3) & 0.48(3) \\
  &  & 0.8 & - & - & 1.60(5) & 0.51(3)\\
\hline
Hinrichsen \cite{hinr01} & 2001 & 0.1 & 0.25 & $<0.67$ & 1.83(5) & 0.50(3) \\
\hline Park, Hinrichsen \& Kim \cite{park01}& 2001 & * & 0.236(10)
& 0.50(5) & 1.80(2) & - \\
\hline Park \& Kim \cite{park02} & 2002 & * & 0.241(5) & 0.496(22)
& 1.80(10) & -\\
 &  & * & 0.242(5) & 0.519(24) &
1.78(5) & -\\
\hline
 Dickman \& de Menezes \cite{dick02} & 2002 & 0.1 & 0.249(5) & 0.546(6) & 2.04(4) & 0.503(6)\\
 &   & 0.5 & 0.236(3) & 0.468(2) & 1.86(2) & 0.430(2) \\
 &   & 0.85& 0.234(5) & 0.454(2) & 1.77(2) & 0.412(2) \\
\hline
 \'Odor \cite{odor03}& 2003 & 0.1 & 0.206(7) & 0.407(7) & 1.95(1) & 0.49(2) \\
 &   & 0.5 & 0.206(7) & 0.402(8) & 1.84(1) & 0.41(2)\\
 &   & 0.7 & 0.214(5) & 0.39(1)  & 1.75(1) & 0.38(2)\\
\hline Kockelkoren \& Chat\'e\cite{kock03}& 2003 & * & 0.200(5) &
0.37(2) & 1.70(5) & - \\
\hline
 Barkema \& Carlon \cite{bark03}& 2003 & 0.1 & 0.17 & - & - & -\\
 &  & 0.2 & 0.17 & - & 1.70(1) & 0.28(4)\\
 &  & 0.5 & 0.17(1) & - & - & 0.27(4)\\
 & & 0.9 & 0.17 & - & 1.61(3) & -\\
\hline
 Noh \& Park \cite{noh04} & 2004 & 0.1 &0.27(4) & 0.65(12) & 1.8(2) & 0.50(5) \\
\hline
 Park \& Park \cite{park05} & 2005 & 1/3 & 0.20(1) & - & - & -\\
\hline Hinrichsen \cite{hinr06} & 2006 & * & $<0.185$ & $<0.34$ &
$<1.65$ & -\\
\hline
 De Oliveira \& Dickman \cite{oliv06} & 2006 & 0.1 & - &
- & 2.08(15) & 0.505(10)\\
 & & 0.5  & - & - & 2.04(5) & 0.385(11) \\
 & & 0.85 & - & - & 1.88(12) & 0.386(5) \\
\hline
 Kwon \& Kim \cite{kwon07} & 2007 & * & - & - & 1.61(1) & -\\

\hline \hline
 Directed Percolation & - & - & 0.1595 & 0.2765 & 1.5807 & 0.2521\\
\hline
\end{tabular}
\caption{ Reported values for the critical exponents of PCPD
\cite{henk04}. In some of the studies, a modified model was used,
changing the definition of the parameters; these studies are
marked with a star in the $d$ column. } \label{tab01}
\end{table*}

\section{Simulation approach}

In our analysis, we use both normal Monte Carlo simulations and
simulations at constant density. For the former, we followed the
usual method of simulating a system of sites that can each contain
a particle, using the same multispin coding program as used by
Barkema and Carlon in an earlier study \cite{bark03}.

It is very hard to get accurate data at low densities (below 10\%)
with standard simulations, due to time constraints and the
possibility of the system reaching an absorbing state because of
its finite size.  Thus, to get more accurate data at lower
densities, we also performed simulations of the PCPD at constant
density. This was done by having two possible reactions for the
system: each step consists of either the usual diffusion, or a
combination of one annihilation of a pair of particles and the
creation of two new ones. It has been shown \cite{wall95,broe99,hilh01} that
this procedure, after thermalization, produces configurations
which are indistinguishable from those obtained with standard
simulations (with a fixed value for $p < p_c$) after the same
density has been reached.

To correctly update the rates of the reactions in this system, the
number of possible places where these reactions can take place is
required. Counting these at each time step would take too much time,
so instead of keeping track of sites, and whether or not they are
occupied, only the existing particles and the distance to their
neighbors are stored during the simulation. Based on the number of
direct neighbors of a particle ($0, 1$ or $2$), the information
is kept in one of three lists. Particles are moved from one list
to another when a reaction causes them to gain or lose a neighbor.
This method allows us to track the number of particles with 0, 1,
and 2 neighbors at every step of the simulation. In addition, it enables
us to pick randomly a particle with a certain number of neighbors, making
sure that we can always perform a chosen reaction. We can just compute the
probabilities for each reaction in a given configuration, choose a reaction
based on that, and pick a site to perform it. This prevents the rejected
reactions normally seen in Monte Carlo simulations, allowing us to
speed up the simulations significantly.

The probability to perform a reaction in a general PCPD system simply
equals the rate (same as in eq.~(\ref{eqrates})) multiplied by the
number of possible places where the reaction can be performed, yielding
\begin{eqnarray}
\begin{array}{ccl}
\left\{
\begin{array}{ccc}
AA0 & \rightarrow & AAA \\
0AA & \rightarrow & AAA
\end{array} \right.
&P_{\mathrm{proc}} &=\,  \frac{(1-p)(1-d)n_1} 2  \\
AA~ \rightarrow ~ 00
&P_{\mathrm{anni}} &=\,  p\,(1-d)(n_1/2 + n_2) \\
A0  \leftrightarrow  0A &P_{\mathrm{diff}} &=\,   d\,(n_0 + n_1 /
2).
\end{array}
\label{eqchance1}
\end{eqnarray}
Here, $n_i$ is the number of particles with $i$ direct neighbours,
and the probability for each reaction is not normalized yet. Since we
want the probability for procreation to be twice as large as that for
annihilation to keep a constant number of particles, we can
compute the value of $p$ that would achieve this from the values of $n_i$
at each simulation step. Knowing this value at each step allows us
to compute the probabilities for each reaction in the model with constant
density using
\begin{eqnarray}
p_{\mathrm{eff}} &=& {n_1 \over {3n_1 + 4n_2}}\label{eqpeff}\nonumber\\
P_{\mathrm{2proc+anni}} &=& p_{\mathrm{eff}}\,(1-d)(n_1/2 + n_2)  \\
P_{\mathrm{diff}} &=&  3 d\,(n_0 + n_1 / 2).\nonumber
\end{eqnarray}
Only two reactions exist now: diffusion, and the combination of
two particle creations and one pair annihilation. Since the latter
reaction is actually a combination of three separate events, the
probability for diffusion is multiplied by a factor of three to
compensate this. The computed value for $p_{\mathrm{eff}}$ can be
monitored over the course of the simulation and averaged to find
the value of $p$ corresponding to the fixed density of the system.

Of course, quantities depending on time cannot be measured from these
simulations at constant density. However, these simulations are very
useful for determining the exponent $\beta$, since it allows us to
explore the relation between $\rho$ and $p$ for low densities much
faster and without the risk that the system reaches an absorbing state
by fluctuations.

Apart from the usual density, the pair density $\rho^*$ was also measured
in both normal and constant-density simulations. This density is simply
the number of pairs of directly adjacent particles, divided by the length
of the system. Since the ratio between $\rho$ and $\rho^*$ approaches
a non-zero constant in the thermodynamic limit, as shown by Barkema and
Carlon \cite{bark03}, $\rho^*$ should obey the same power laws as $\rho$,
with the same exponents. Finite-size or finite-time effects might however
not be exactly the same for $\rho$ and $\rho^*$. Therefore, if we find
different critical exponents for the particle-density and pair-density,
we know that the method of analysis used is incorrect.

\section{Direct computation of critical exponents}

\begin{figure}
\includegraphics[width=7.7cm]{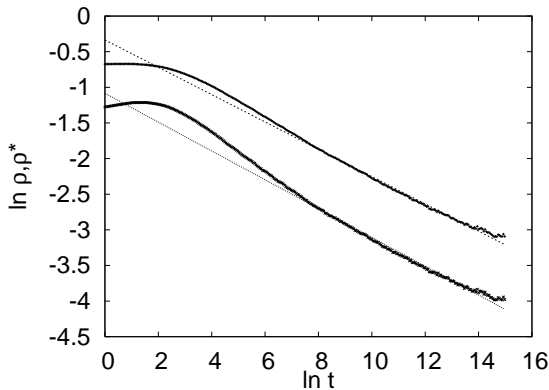}
\caption{Particle density $\rho$ and pair density $\rho^*$ as a
function of time, for PCPD simulations at $d=0.5$, $p=0.1524$ and
$L=100,000$, averaged over 64 simulations. The straight line is a
fit to determine $\delta$ ignoring correction terms. The top curve
shows the density, with $\delta_{\mathrm{direct}} = 0.19$. The bottom
curve shows the pair density, with $\delta_{\mathrm{direct}} =
0.20$.} \label{NDeltaPlot}
\end{figure}

A direct analysis of the simulation data from all performed simulations
is fairly straightforward, and has already been shown before to give rise
to great inaccuracies in its results.  In this section we will give a
quick overview of this direct analysis, with particular attention for
the differences between the results for the particle density $\rho$
and pair density $\rho^*$.  A first estimate of the critical exponent
$\delta$ is obtained by taking the logarithm of both the density and time
in a simulation close to the critical point, and fitting a straight line
through the data. The simulations used to determine $\delta$ were run on a
system with $L=100~000$, for about $3\cdot 10^6$ timesteps. For $d=0.5$,
this leads to $\delta=0.19$, as shown in Figure \ref{NDeltaPlot}. A
slightly higher value for $\delta$ is found, however, if this analysis
is performed on the pair density instead ($\delta^* = 0.20$), and it
is visible that the double-logarithmic curve is not entirely straight.
As was already shown before \cite{bark03, hinr06}, this curving tendency
can very well be extrapolated to $\delta_{\mathrm{DP}}$.

\begin{figure}
\includegraphics[width=7.7cm]{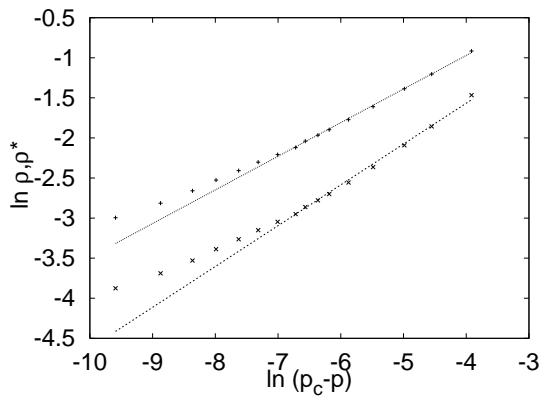}
\caption{The particle density $\rho$ and pair density $\rho^*$ as
a function of $p-p_c$ in constant-density simulations, with
$d=0.5$ and $L=100,000$. The data for the pair density is the
lower set. The fits, using only the points with higher densities,
yield $\beta_{\mathrm{direct}}=0.42$ for the particle density, and
$\beta_{\mathrm{direct}}=0.51$ for the pair density.}
\label{NBetaPlot}
\end{figure}
\begin{figure}
\includegraphics[width=7.7cm]{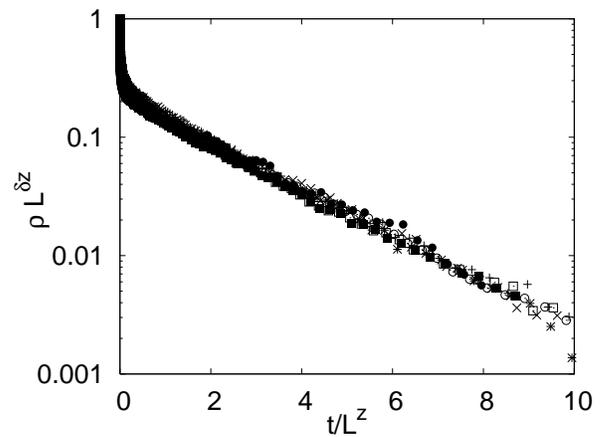}
\caption{ Data collapse to determine $z$, ignoring correction
terms. Each curve is the average over 3200 simulations, for
$d=0.5$, $p=0.1524$, and $L=200 \, (+)$, $300 \,(\times)$, $500
\,(\ast)$, $1000 \,(\boxdot)$, $2000 \,(\blacksquare)$, $3000
\,(\odot)$, and $5000 \,(\bullet)$. we find $z_{\mathrm{direct}} =
1.75$. Since the particle density of the system will not always tend
to $0$ due to the possibility of a single remaining particle, we only
include the data on the pair density.} \label{NZPlot}
\end{figure}

The value of $\beta$ can be determined by performing simulations
at non-critical values of $p$, where the system reaches its steady
state. Here, we used simulations at constant density for this.
The exponent $\beta$ can be extracted from a logarithmic plot of
the density versus $p-p_c$. Figure \ref{NBetaPlot} shows such a
plot. At higher densities, this yields values for $\beta$ which are
consistent with earlier reported measurements, significantly higher than
$\beta_{\mathrm{DP}}$. However, it is clear that this fit does not hold
up for the lower densities, which can be reached in the constant-density
simulations. In addition, for the region with higher densities, the slopes
for the particle and the pair density are clearly different, showing
that an analysis consisting of a simple linear fit cannot be trusted.

For the calculation of the critical exponent $z=\nu_\parallel/\nu_\perp$
we examine finite-size effects on a system at the critical value of
$p$. If the system size $L$ is small, these effects will cause the
density to start decaying exponentially once the correlation length
$\xi \sim t^{1/z}$ approaches the system size. Since the particle and
pair densities are tied closely together, they will collapse at the same
time. After sufficient time, the density will then decay as:

\begin{eqnarray}
\rho &\sim & \exp(-b({t/L^z})) \Leftrightarrow\\
\nonumber\\
\ln (\rho)&=& a-b (t/L^z).
\label{zcollapse}
\end{eqnarray}

Using Eq.~(\ref{zcollapse}), the exponent $z$ can be obtained from
simulations in small systems of various sizes until past this point of
collapse, and adjusting the exponent $z$ until a data collapse can be
attained for the exponential regime.  We simulated systems for $d=0.1,
0.5$ and 0.9, with system sizes ranging from $L=200$ to 5000 sites. For
system sizes larger than this, the collapse occurs at late times, and
therefore at a very low density. Apart from the very long simulation
times, this poses another problem. To have accurate data at that point,
we would have to know $p_c$ accurately enough such that its error
$\Delta p_c$ obeys $(\Delta p_c)^\beta \ll \rho$. With the density
dropping below $0.04$ near the point of collapse for larger systems,
$\Delta p_c$ would have to be smaller than $10^{-5}$ to avoid significant
systematical errors. Since our precision in $p_c$ is not that accurate,
our range for the system size is limited by this effect.

At the time the density starts collapsing, $t \sim L^z$. Since up to this
point, the density was following a power law, we know that the density
at this point will be $\rho_{coll} \sim t^{-\delta} \sim L^{-\delta z}$.
Using this, we can scale the data from our simulations to obtain a data
collapse, as shown in Figure \ref{NZPlot} for $d=0.5$. It turns out that
scaling the vertical axis works best with $\delta z$ = $\delta_{DP} z_{DP}$,
though the optimal value for $z$ in the horizontal scaling again varies
with $d$.

Table \ref{NTable} shows the results for all exponents, for the
three values of $d$ we investigated. The exponents all vary when
the diffusion parameter $d$ changes, which, assuming PCPD falls
into a single universality class, again points out there is
something wrong with such a direct analysis.

\begin{table}
\begin{tabular}{|c||c|c|c|c|c|c|}
\hline $d$ & $p_c$ & $\delta$ & $\delta^*$ & $\beta$ & $\beta^*$& $z$ \\
\hline
0.1 & 0.1111 & 0.22 & 0.24 & 0.48 & 0.60 & 1.83\\
0.5 & 0.1524 & 0.19 & 0.20 & 0.42 & 0.51 & 1.75\\
0.9 & 0.2333 & 0.19 & 0.20 & 0.34 & 0.39 & 1.64\\
\hline
\end{tabular}
\caption{The results of the direct analysis of the critical
exponents, ignoring correction terms.}
\label{NTable}
\end{table}

\section{Analysis including corrections}

\begin{figure}
\includegraphics[width=7.7cm]{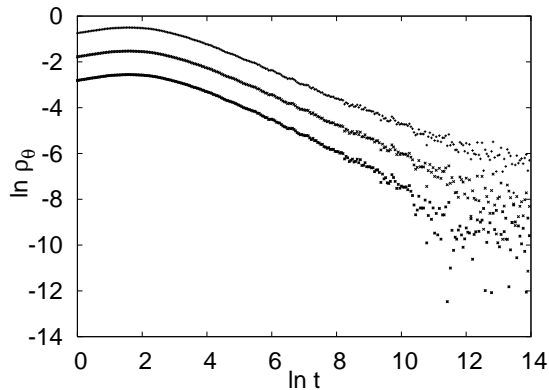}
\caption{ The decay in time of linear combinations of the particle
and pair densities, plotted to find the combination that isolates
the correction term. Here, $\rho_\theta = a/a^* \rho^* - \rho$,
with $a/a^*=2.41,\, 2.43$ and $2.45$ (from bottom to top), at $d =
0.5$ and $p=0.1524$. The top and bottom lines are shifted up and
down by 1 for clarity. The middle plot is straightest; from its
slope we get the correction exponent $\theta=0.63$.}
\label{ThetaPlot}
\end{figure}

Given the problems with the direct analysis of the critical
exponents $\delta$ and $\beta$, we propose introducing correction
terms into our scaling laws. Both $\rho$ and $\rho^*$ obey the
power laws as before, but with the DP values for the leading term
in each relation, and an added correction term with a higher
exponent. For the time-dependence of the density, this yields
\begin{eqnarray}
\begin{array}{rcl}
\rho & = & a t^{-\delta} + b t^{-\theta} \\
\rho^* & = & a^* t^{-\delta} + b^* t^{-\theta}.
\end{array}\label{delteq1}
\end{eqnarray}
While it is expected that the exponents for the two relations are
the same, the prefactors for the particle density can be different from
those for the pair density. This makes it possible to determine
linear combinations of $\rho$ and $\rho^*$ where one of the two
terms is suppressed, using
\begin{eqnarray}
\begin{array}{rcl}
\rho_\delta \equiv \rho - {b \over b^*} \rho^* & = & c t^{-\delta} \\
\rho_\theta \equiv {a \over a^*} \rho^* - \rho & = & \tilde{c}
t^{-\theta}.
\end{array}\label{delteq2}
\end{eqnarray}
To determine all coefficients accurately, we will need both exponents. The
correction exponent $\theta$ can be computed once a value for $a/a^*$
has been found that turns $\ln \rho_\theta$ into a linear function of
$\ln t$, at least for low densities. As a starting estimate for this
ratio, an extrapolation of the ratio $\rho/\rho^*$ for $\rho \to 0$
can be used. Figure \ref{ThetaPlot} shows the plots used to determine
$a/a^*$ and calculate $\theta$.

\begin{figure}
\includegraphics[width=7.7cm]{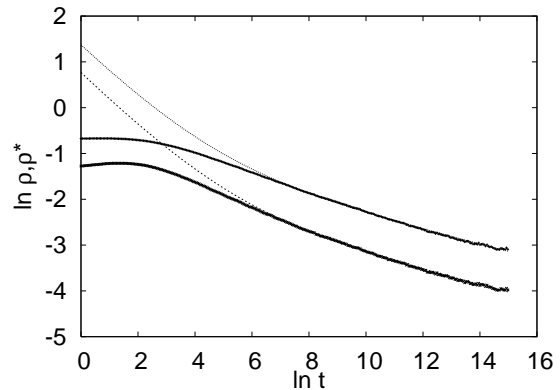}
\caption{ Time-dependance of the particle density $\rho$ (top) and pair
density $\rho^*$ (bottom) at $d=0.5$, $p=0.1524$ and $L=100,000$, averaged
over 64 simulations.  The lines were obtained from a least-squares
fit of both $t^{-\delta}$ and the correction term $t^{-\theta}$ to
the data, using $\delta=\delta_{\mathrm{DP}}$ and $\theta=0.63$.}
\label{RhoFitPlot}
\end{figure}

With the exponent $\theta$ known, and using the DP value for
$\delta$, it is now possible to fit both $\rho$ and $\rho^*$ with a
linear combination of $t^{-\delta}$ and $t^{-\theta}$, to obtain the
prefactors $a, b, a^*$ and $b^*$. These fits are shown for $d=0.5$
in Figure \ref{RhoFitPlot}. The consistency of our fits can then be
checked by determining $\delta$ and $\theta$ again from a linear fit
to logarithmic plots of the appropriate linear combinations of $\rho$
and $\rho^*$. In addition, the assumption for $a/a^*$ can be checked to
make sure that it equals the ratio between the values from the fit.

\begin{figure}
\includegraphics[width=7.7cm]{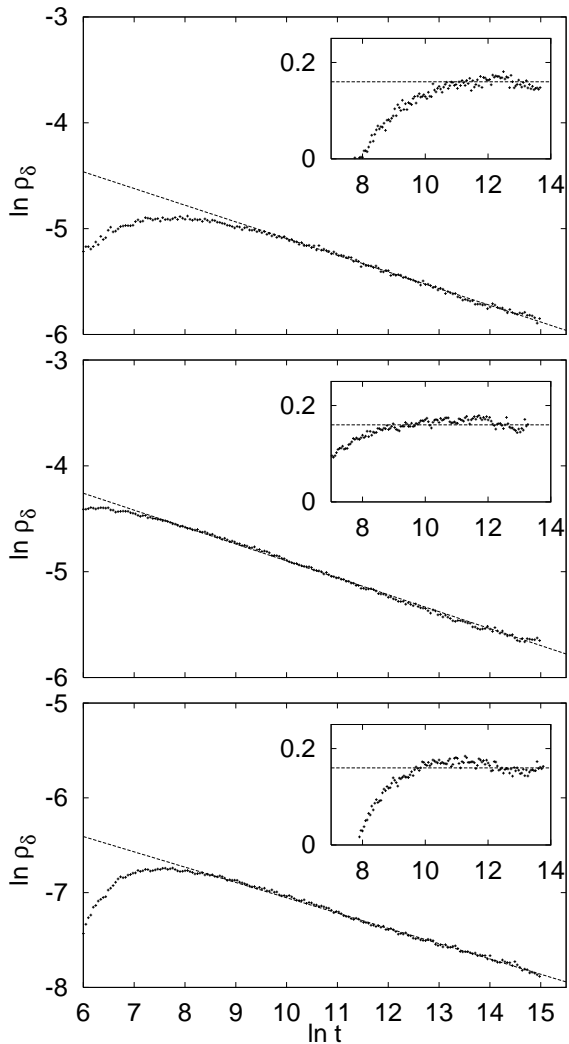}
\caption{ Fit to determine $\delta$ from $\rho_\delta$, the ideal
linear combination of the particle and pair density that
suppresses the correction term. The slope of the fit yields
$\delta=0.16(1)$, at (from the top graph to the bottom) $d=0.1$
and $p=0.1111$, $d=0.5$ and $p=0.1524$, and $d=0.9$ and
$p=0.2333$, with all data averaged over 64 simulations at each
diffusion rate. The insets show the effective exponent
$\delta_{\mathrm{eff}} = \partial \ln (\rho_\delta) / \partial
\ln t$ as a function of $\ln t$, with the horizontal line at
$\delta = 0.16$. } \label{DeltaPlot}
\end{figure}

Going through this process for $d=0.5$, the final fit to determine
$\delta$ from the computed ideal combination of $\rho$ and $\rho^*$ gives
a value of 0.16(2), as shown in Figure \ref{DeltaPlot}, consistent with
$\delta_{\mathrm{DP}} = 0.1595$. The Figure also shows the effective
exponent as it changes during the simulation. The remaining curvature
in these graphs is sensitive to small changes in the estimation
of $a/a^*$, even if those do not significantly affect the resulting
exponents. Therefore, this deviation from a straight line is most likely
caused by an inaccuracy in this estimation.  The differences between
the estimated and calculated ratio $a/a^*$ and between the two obtained
values for $\theta$ are well within the error margins for those values. In
Table \ref{DeltaTable}, the results for the fit are shown also for other
values of $d$.
\begin{table}
\begin{tabular}{|c||c|c|c|c|c|c|c|c|c|}
\hline
$d$ & $a/a^*$ & $\theta$ &  a   & $b$  & $a^*$ & $b^*$ & $\delta$ & $\theta_{\mathrm{check}}$ \\
\hline
0.1 & 2.56    & 0.531    & 0.41 & 3.60 & 0.16  & 2.44  & 0.16(2) & 0.571  \\
0.5 & 2.43    & 0.625    & 0.47 & 3.47 & 0.19  & 1.96  & 0.16(2) & 0.633  \\
0.9 & 4.19    & 0.858    & 0.37 & 27.0 & 0.089 & 9.31  & 0.16(2) & 0.887  \\
\hline
\end{tabular}
\caption{The results of our analysis for $\delta$, if a correction term
is included. The $\theta_{\mathrm{check}}$ column shows the value of
the exponent $\theta$ obtained by using $a$ and $a^*$ from the fit,
to cross-check the values obtained in Figure \ref{ThetaPlot}.}
\label{DeltaTable}
\end{table}

As seen in the table, the exponent $\theta$ seems to vary as $d$
changes. It is likely, however, that this is the effect of further
correction exponents, whose coefficients depend on $d$.  Given the
numerical precision of our data, it would be too optimistic to
claim that our values for $\theta$ are accurate. A fit of Eq. (7)
with $\delta=0.1595$ to the high-quality data of Ref. \cite{hinr06},
which runs up to time $t=10^8$, yields correction exponents as low as
$\theta=0.3$. This shows that either the corrections to $\theta$ are very
strong, or that the leading finite-time corrections actually cancel out in
$\rho_\theta$, causing us to measure the next correction exponent instead.

A similar process can be followed for determining $\beta$, using
the constant density simulations. Again, we used $L=100,000$, with
simulation times varying based on the relaxation time of the
system. Our densities range from $0.05$ to $0.4$. For low
densities, it can take up to $10^9$ simulation steps until the
system no longer shows a systematic drift. The assumed behavior of
the densities is
\begin{eqnarray}
\begin{array}{rcl}
\rho & = & a (p_c-p)^{\beta} + b (p_c-p)^{\zeta} \\
\rho^* & = & a^* (p_c-p)^{\beta} + b^* (p_c-p)^{\zeta}\\
\rho_\beta & \equiv & \rho - {b \over b^*} \rho^*  =  c (p_c-p)^{\beta} \\
\rho_\zeta & \equiv & {a \over a^*} \rho^* - \rho  =  \tilde{c}
(p_c-p)^{\zeta}.
\end{array}
\end{eqnarray}
Of course, the values for the prefactors, as well as the
correction exponent, will be different from those in equations
\ref{delteq1}. However, the value for $a/a^*$ should still be the
thermodynamic limit of the ratio $\rho/\rho^*$, and thus will be
the same as in our calculation of $\delta$.

\begin{figure}
\includegraphics[width=7.7cm]{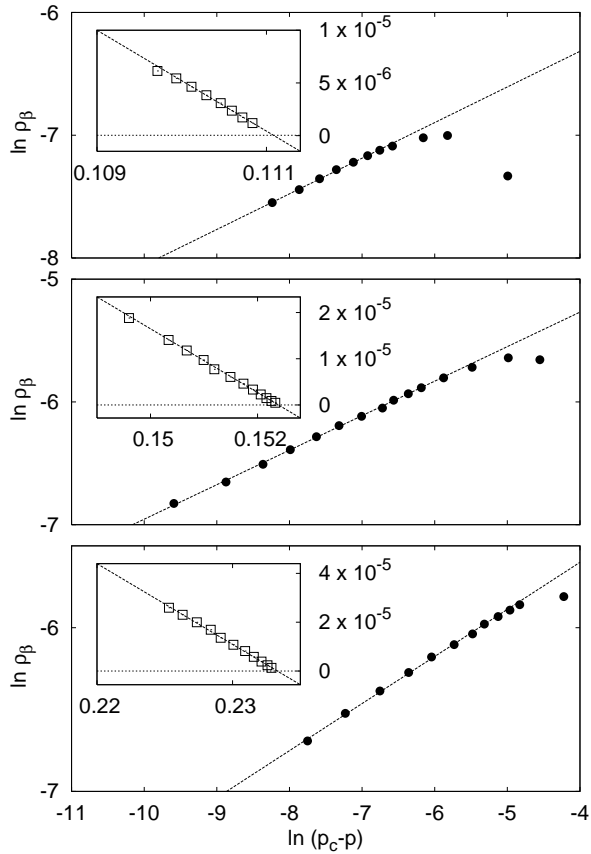}
\caption{ Plots of the linear combination $\rho_\beta$ of the
particle density and pair density that suppresses correction terms
in the calculation of $\beta$, using constant-density simulations
at $d=0.1$ (top), $d=0.5$ (middle) and $d=0.9$ (bottom), at
$L=100,000$. The slope of the line gives $\beta = 0.29(3),
0.28(2)$ and $0.29(3)$, from top to bottom, in agreement with DP.
The insets show how $p_c$ for each value of $d$ was checked using
these simulations, by plotting the same $(\rho_\beta)^{1/\beta}$
as a function of $p$.}\label{BetaPlot}
\end{figure}

Plotting the linear combination of $\rho$ and $\rho^*$ which
suppresses the leading term, using the same ratio of $a/a^*$ as
before, we can determine the first correction exponent $\zeta$.
With the exponents of the two leading terms known, we can again
fit these to the data to determine their prefactors, and find the
linear combination which will suppress the correction term. The
result of this is shown in figure \ref{BetaPlot}. Again, the
values from the fit to cross-check $a/a^*$ and $\zeta$ show that these
estimates are consistent with the fit. Table \ref{BetaTable} shows
the results for different values of $d$. Again, the correction
exponent varies slightly as the diffusion parameter $d$ changes,
suggesting additional corrections beyond the first. However, all
data is consistent with a $\beta_{\mathrm{DP}} = 0.2765$.

To confirm that we are performing our simulations at the critical
point, we can calculate $p_c$ from these constant density
simulations as well. With the same linear combination of the
density and pair density as used for calculating $\beta$, we again
suppress the correction term. Since $(\rho_\beta)^{1/\beta}$ is a
linear function of $p$, and equals 0 at $p=p_c$, finding $p_c$ is
a matter of linear extrapolation, as shown in the insets in Figure
\ref{BetaPlot}. The results are in agreement with the values for
$p_c$ used for all our simulations at the critical point, as seen
in Table \ref{BetaTable}.
\begin{table}
\begin{tabular}{|c||c|c|c|c|c|c|c|c|}
\hline
$d$ & $\zeta$ & a & $b$ & $a^*$ & $b^*$ & $\beta$ & $\zeta_{\mathrm{check}}$& $p_c$ \\
\hline
0.1 & 1.058 & 0.62 & 43.1 & 0.24 & 26.5 & 0.29(3) & 1.055 & 0.1111 \\
0.5 & 1.056 & 0.70 & 16.4 & 0.29 & 10.5 & 0.28(2) & 1.070 & 0.1524 \\
0.9 & 1.270 & 0.43 & 17.0 & 0.10 & 7.77 & 0.29(3) & 1.285 & 0.2333 \\
\hline
\end{tabular}
\caption{The results of our analysis for $\beta$, when including a
correction term. The $\zeta_{\mathrm{check}}$ column shows the
value of the exponent $\zeta$ obtained by using $a$ and $a^*$ from
the fit, which allows us to confirm our calculation of $\zeta$.}
\label{BetaTable}
\end{table}

Lastly, we turn to the exponent $z$. While the data collapse in
figure \ref{NZPlot} is acceptable, we do find different values of
$z_{\mathrm{direct}}$ at $d=0.1, 0.5$ and $0.9$. Since we have
demonstrated that equal values for the exponents $\beta$ and
$\delta$ can be obtained for different values of $d$ by including
correction terms, this encourages us to try out a similar approach
here.

As before, we add a correction term to the relevant equation, in this
case the one for the density during the collapse, and see how this
fits with the DP exponents. The density will now decay differently,
following: 
\begin{equation}
\rho_{coll}  \sim  \exp (-b t(L^{-z} + c L^{-\chi})),
\end{equation}
with $b$ an unknown constant and $\chi$ the correction exponent, with
prefactor $c$. Considering that one of the determining factors of the
system is diffusion, where the correlation length grows as $\sqrt{t}$,
a correction exponent of $\chi=2$ is a reasonable assumption. Sadly, with
the range of data available it is impossible to determine this exponent
accurately.  We can, however, assume $z=z_{\mathrm{DP}}$ and $\chi=2$,
and show that this will provide data collapses at least as acceptable as
the ones without a correction term. We use the same scaling as before,
though we include the corrections calculated in our analysis for $\delta$
in our vertical scaling: 
\begin{equation}
\rho_{coll} = \rho^*(t=L^z) \sim L^{\delta z}
+ {b^*\over a^*} L^{\theta z},
\end{equation}
with $\theta$, $a^*$ and $b^*$ taken from Table \ref{DeltaTable}.

\begin{figure}
\includegraphics[width=7.7cm]{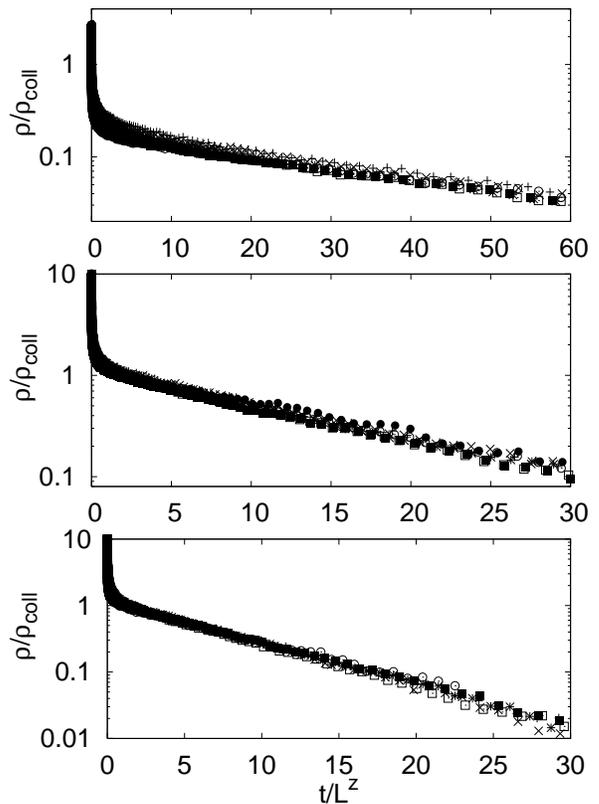}
\caption{ Data collapses with $z=z_{DP}$ for $d=0.1,p=0.1111$
(top), $d=0.5, p=0.1524$ (middle) and $d=0.9,p=0.2333$ (bottom),
at system sizes $L=200 (+)$, $300 (\times)$, $500 (\ast)$, $1000
(\boxdot)$, $2000 (\blacksquare)$, $3000 (\odot)$ and 5000
($\bullet$). All data is averaged over 3200 runs. The values for $c$
used are $c=20, 15$, and $3$, in order of increasing $d$.}
\label{ZPlot}
\end{figure}

Note that there still is only one free parameter in our data
collapse after these assumptions have been made: instead of
varying the exponent $z$, modifying the prefactor $c$ is now the only
way to make the data from different system sizes overlap, both for
the particle density and the pair density. This prefactor for
the correction term can vary as $d$ changes, however. By plotting
this relation for different sizes, we can find the value for $c$
that causes a data collapse, as well as the equivalents for the
pair density. The ratio of the particle density and pair density
varies very little in the regime we are examining here, and the
point of collapse is equal for both, so we expect that $b$ and
$b^*$ are equal. Fig. \ref{ZPlot} shows the data collapses for
three values of $d$. While this does show that all of our data can
be seen as consistent with $z_{\mathrm{DP}}$, there is no clear
way to show that this interpretation is better than the simpler
approach, which yields varying exponents for different values of
$d$. However, it does seem more likely that if the other two
exponents are independent of the diffusion rate, the same should
hold for $z$.

\section{Conclusion}
Monte Carlo simulations, both the usual approach with constant rate $p$
and a new approach at constant density, have been used to analyze the
critical behavior of the pair contact process with diffusion. While
many recent studies conclude that this model does not belong to the
directed percolation universality class, we find that if correction
terms are included in the power laws governing critical scaling, all of
the acquired simulation data is consistent with the exponents from DP.

In our simulations, especially the calculations for $\beta$ and $\delta$
offer convincing evidence that the DP values for these critical exponents
are indeed accurate for the PCPD model. Since the critical exponents
for $\rho$ and $\rho^*$ must be equal, any linear combination of these
must have the same exponents as well, in the thermodynamical limit,
except in the singular case where the leading terms cancel out. The fact
that there exists a linear combination which follows a power law that is
consistent with the DP exponent shows that the correct exponent for the
system can at least not be any larger than that. Our analysis of $z$
does not lend itself to an accurate calculation of the exponent, but
does also show that a correction term can explain the deviation from the
DP exponent.  With all of our simulation data consistent with DP for all
three of the studied exponents, and for all investigated diffusion rates,
we conclude it is likely that PCPD does belong to the directed
percolation universality class.

We thank Haye Hinrichsen for providing us with the data described in Ref. \cite{hinr06}.

\end{document}